# GenCos' Behaviors Modeling Based on Q Learning Improved by Dichotomy

Qiangang Jia, Zhaoyu Hu, Yiyan Li,
Zheng Yan and Sijie Chen
Key Laboratory of Control of Power
Transmission and Conversion, Ministry
of Education
Shanghai Jiao Tong University
Shanghai, 200240 China
jiaqiangang@sjtu.edu.cn

*Abstract*—Q learning is widely used to simulate the behaviors of generation companies (GenCos) in an electricity market. However, existing Q learning method usually requires numerous iterations to converge, which is time-consuming and inefficient in practice. To enhance the calculation efficiency, a novel Q learning algorithm improved by dichotomy is proposed in this paper. This method modifies the update process of the Q table by dichotomizing the state space and the action space step by step. Simulation results in a repeated Cournot game show the effectiveness of the proposed algorithm.

## I. Introduction

Reinforcement learning [1], supervised learning, and unsupervised learning are the basis of current artificial intelligence. As a method with self-learning ability, reinforcement learning can be further divided into model-based learning and model-free learning [2]. In model-free learning, Q learning [3] is a typical method based on temporal difference[4], which is a combination of the Monte Carlo [5] method and dynamic programming. Q learning is usually applied to incomplete information games to mimic participants' behaviors, which provides a valuable insight for operators to evaluate the rules of the games.

The power market represents a typical incomplete information game for GenCos. Typically, several GenCos can exercise market power to manipulate the price. Many researchers have applied the Q learning algorithm to model GenCos' behaviors, which helps market operators assess the rationality of current market rules, e.g., in mitigating market power. Yu introduced the Q learning method to evaluate the market rules in the California electricity market [6]. Kebriaei used the fuzzy Q learning to mimic GenCos' bidding strategies in a Cournot oligopoly electricity market [7]. Note that the Q learning algorithms in the above papers need to discretize the state space and action space with a certain step size. The smaller the size is, the higher the precision is, but the lower the calculation efficiency is. Also, a high granularity of the discretization may introduce the curse of dimensionality and slow down the market rule evaluation. Deep learning [8] has been studied to solve the problem by transforming discrete state spaces into continuous state spaces, but the deep neural network needs massive data to train, which is infeasible in such a market. Thus, finding a proper method to enhance the calculation efficiency is the key to improve the performance of Q learning in the modeling of GenCos.

This paper brings the idea of dichotomy into Q learning to reduce the number of iterations. Based on dichotomy, both the state space and the action space can be dichotomized step by step, rather than be discretized arbitrarily. To verify its effectiveness, we choose a repeated Cournot game [9] as the environment to implement the simulation. Simulation results show its effectiveness in reducing the number of iterations.

The rest of the paper is organized as follows. Section II introduces the framework of the Q learning algorithm improved by dichotomy. Section III details the proposed algorithm in a repeated Cournot game. Case studies are presented in Section IV. Finally, Section V gives the conclusion and future research direction.

## II. Q Learning Improved by Dichotomy

### A. Q Learning Process Based on Q Table

The Q learning process is based on the Q table whose rows and columns represent the state spaces and the action spaces respectively. The Q value of each state-action pair is saved in the Q table. The whole process of Q learning can be decomposed into the following four steps.

1) Initialize the Q table.

2) Select an action based on the current state according to a specific policy, including Softmax policy [10] and ε-greedy policy, to balance the exploration and the exploitation.

3) Obtain rewards and turn to the next state.

4) Update the Q table. Then turn to step 2) if the Q table does not converge. The update formula of the Q table based on temporal difference is given by

$$Q_{t+1}(x,a) = Q_t(x,a) + \alpha(r + \gamma \max_{a'} Q_t(x',a') - Q_t(x,a)) \quad (1)$$

where $t$ denotes the iteration times, $x$ denotes the current state, $a$ denotes the current action, $r$ denotes the reward, $\alpha$ denotes the learning rate, $\gamma$ denotes the discount factor, $x'$ denotes the next state, and $a'$ denotes the action to maximize the Q value.

### B. Dichotomy

Dichotomy is a method to dichotomize the searching zone step by step, which is widely used to search for items among bulk data in computer science.

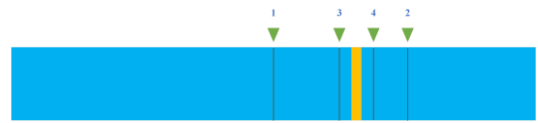

Fig. 1. The basic idea of dichotomy

In Fig.1, the blue part and the yellow part denote the whole searching area and the target respectively, the green triangles and black lines denote the positions of dichotomy. The main idea is continuously dichotomizing the searching area to locate the target. Generally, the time complexity of the algorithm is $O(logn)$ when the searching area is linear.

*C. Combinations of Q Learning and Dichotomy*

The most time-consuming part in Q learning is the update process of the Q table, which will be accelerated greatly by introducing dichotomy. The flow chart is shown in Fig.2.

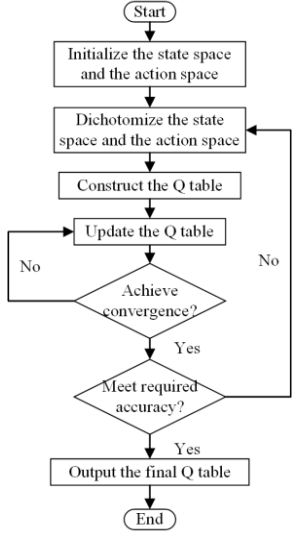

Fig. 2. The framework of dichotomy in Q learning

The whole learning process is modified based on the above process.

### III. GENCOs' Modeling Based on the Game Theory

In this paper, we choose the Cournot game, which is the most representative model, to construct the oligopoly market in the power supply side.

*A. Q Learning Under Repeated Cournot Games*

Cournot game is also known as Cournot duopoly game, which describes an oligopoly market in which two independent players compete with each other by changing their supply quantities simultaneously. Generally, the game exists a balance called Nash equilibrium (NE) [11], which means nobody can improve its profit by changing behaviors unilaterally.

Repeated Cournot game refers to a Cournot game that is played repeatedly by the same players. In a repeated Cournot game, each GenCo should decide its optimal action according to its competitor's action. The competitor's action is a predicted value in the real power market. However, the prediction mentioned above does not affect the enhancement of calculation efficiency. Thus we assume that the competitor's next action is the same as the previous. Moreover, $\gamma$ is set to zero because the state transitions do not affect the value function. Hence the update formula of the Q table is simplified from (1) to (2) as follows.

$$Q_{t+1}(x_{es},a) = Q_t(x_{es},a) + \alpha(r - Q_t(x_{es},a)) \quad (2)$$

where $x_{es}$ denotes the estimated action of the competitor. Moreover, we choose the Softmax policy to make a balance of exploration and exploitation.

*B. Market Environment*

The market-clearing price (MCP) [12] in the Cournot game is shown via the following formula.

$$p_{clear} = \lambda - \alpha(P_1 + P_2) \quad (3)$$

where $p_{clear}$ denotes the MCP, $\lambda$ denotes the upper limit of MCP, $\alpha$ denotes the slope, $P_1$ and $P_2$ denote the bidding quantities of each GenCo.

*C. GenCos Model*

Each GenCo's cost can be written in a quadratic form that is widely used is the power market [13].

$$C_i = a_i^2 P_i + b_i P_i + c_i \quad (4)$$

where $C_i$ denotes the total cost of GenCo $i$, $a_i$ denotes the coefficient of the quadratic term, $b_i$ denotes the coefficient of the primary term, $c_i$ denotes the constant term, $P_i$ denotes the bidding quantities. The profit function of a GenCo is given by

$$R_i = P_i(\lambda - \alpha\sum_{j=1}^{2} P_j) - a_i^2 P_i - b_i P_i - c_i \quad (5)$$

where $R_i$ denotes the profit of the GenCo $i$ after market clearing. The GenCos would calculate their optimal bidding quantities analytically if they knew all the parameters clearly, which is impossible in reality. Thus a proper learning method interacting with the unknown environment is necessary to help the participants make optimal decisions.

*D. Improved Q Learning Process*

The three most essential elements in Q learning are the state space $S$, the action space $A$, and the reward $R$. For each GenCo, the three elements correspond to the competitor's bidding quantities, own bidding quantities, and the reward $R_i$, respectively. Assuming that two independent GenCos labeled by $x$ and $y$ know each other's generation caps, which are $P_{xmax}$ and $P_{ymax}$ respectively.

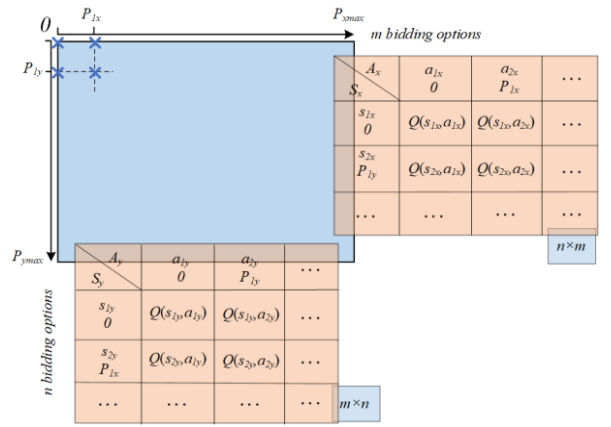

Fig. 3. The traditional Q table

Traditionally, we should construct a Q table as Fig.3. GenCo $x$ has $m$ bidding options while GenCo $y$ has $n$ bidding options. The GenCo's bidding options can be regarded as its action space. Meanwhile, each GenCo's action space is its competitor's state space. Therefore, the dimensions of the two Q tables are $n \times m$ and $m \times n$ respectively. But the high granularity of the discretization may reduce calculation efficiency seriously. Thus the algorithm should be modified.

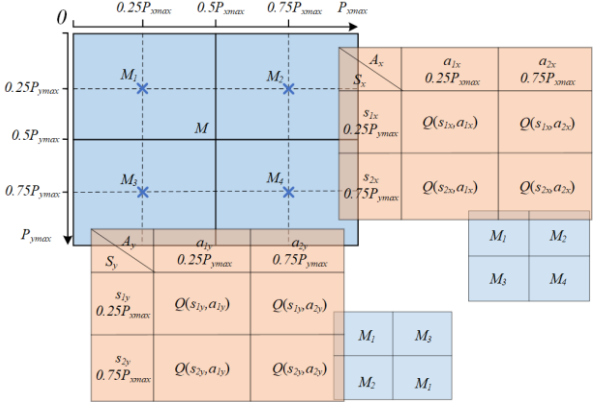

Fig. 4. The improved Q table by introducing dichotomy

Fig.4 shows the basic idea of the algorithm, which consists of several iteration rounds. In the first iteration round, each GenCo dichotomizes the initial variation range of the bidding quantities. Thus the main area centered on $M$ will be sliced into four subareas centered on four different points labeled by $M_1$, $M_2$, $M_3$, and $M_4$, which correspond to different positions in the Q tables of $x$ and $y$. Then each GenCo begins to update the Q table through serval iterations. The current iteration round will be finished if any point in the Q table is larger than a specific number, which is 0.9 in this paper.

- If $Q(s_{1x}, a_{1x})$ as well as $Q(s_{1y}, a_{1y})$ is larger than 0.9, the searching area will be contracted to the rectangle centered on $M_1$.
- If $Q(s_{1x}, a_{2x})$ as well as $Q(s_{2y}, a_{1y})$ is larger than 0.9, the searching area will be contracted to the rectangle centered on $M_2$.
- If $Q(s_{2x}, a_{1x})$ as well as $Q(s_{1y}, a_{2y})$ is larger than 0.9, the searching area will be contracted to the rectangle centered on $M_3$.
- If $Q(s_{2x}, a_{2x})$ as well as $Q(s_{2y}, a_{2y})$ is larger than 0.9, the searching area will be contracted to the rectangle centered on $M_4$.

Then each GenCo will update the variation range of the bidding quantities according to the two adjacent sides of the rectangle centered on $M_i$, which is equal to $M$ in the previous iteration round. Also, each GenCo dichotomizes its variation range of bidding quantities for a new iteration round. After several iteration rounds, a precise NE point will be found out.

## IV. SIMULATION

The simulation is run at Python 3.6 on a PC with an Intel Core i7 CPU and 16GB RAM.

### A. Parameters Setting

The parameters are set in TALBE Ⅰ. For simplicity, the symbols here omit units.

TABLE I. PARAMETERS SETTING

| GenCos | Parameters | | | | | |
|---|---|---|---|---|---|---|
| | $\lambda$ | $\alpha$ | $a$ | $b$ | $c$ | $P_{max}$ |
| $x$ | 102 $\pm 2$ | 0.04 | 0.001 | 2 | 10000 | 2000 |
| $y$ | | | 0.002 | 3 | 11000 | 1800 |

To make the curves smoother rather than stepped, four different parameter sets with different values of $\lambda$ are randomly generated and each parameter set is simulated 20 times to obtain the average. In the traditional Q learning, the action space of $x$ is {0, 50, 100, …, 2000}, the action space of $y$ is {0, 50, …, 1800}. In the improved Q learning, the initial variation range of bidding quantities of $x$ and $y$ are [0,2000] and [0,1800] respectively.

### B. Simulation Results

This paper compares the mean reward, the action and the mean accumulative reward of the traditional method and the improved method. The simulation results are as follows.

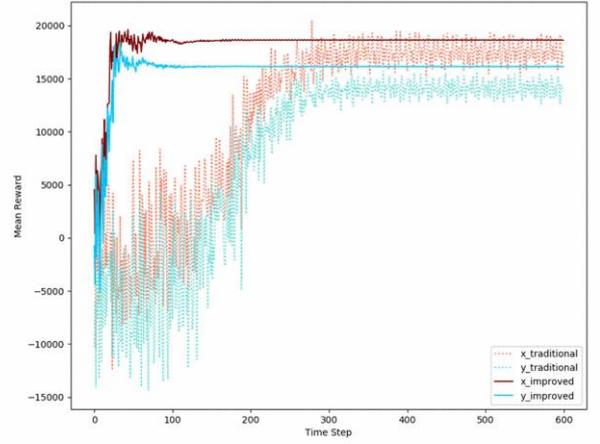

Fig. 5. The mean rewards

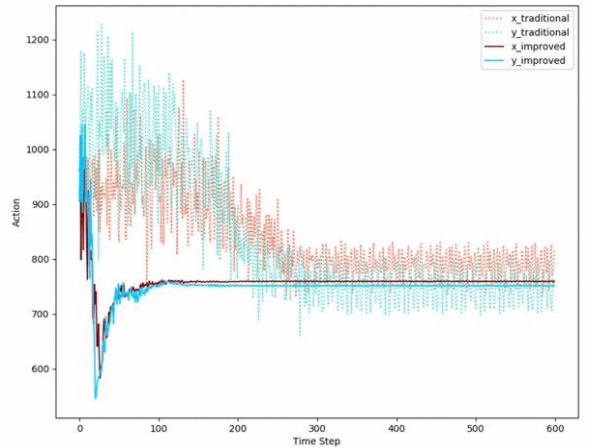

Fig. 6. The actions

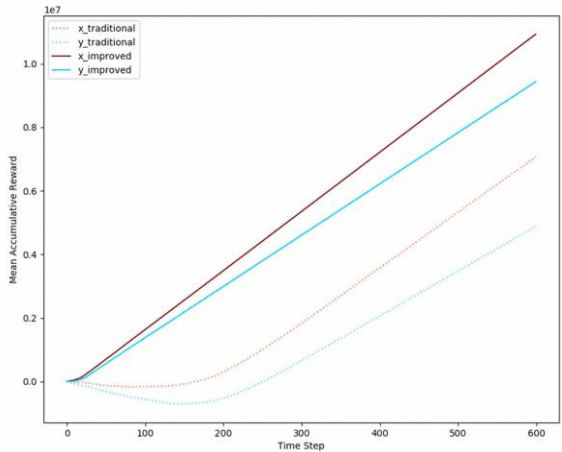

Fig. 7. The mean accumulative rewards

In Fig.5 and Fig.6, we can find that the mean rewards and actions of *x* and *y* still fluctuate weakly after the 300th iteration in the traditional method. However, the improved method achieves convergence in the 100th iteration. Fig.7 shows that *x* and *y* begin to make profits in the 155th iteration by the traditional method while the improved method starts to be profitable in the 22nd iteration.

Simulation results show that the improved Q learning algorithm has a better performance in the calculation efficiency. Besides, the improved method gains a higher profit than the traditional method, which means that the improved method finds a more precise NE point.

## V. CONCLUSION

This paper proposed a novel method to update the Q table in Q learning process by dichotomizing state spaces and action spaces step by step. Simulation results show its effectiveness in enhancing the calculation efficiency. This algorithm will help the market operators evaluate and improve the electricity market rules more effectively. Further study will concentrate on its applications in a more complex environment with multiple participants.

REFERENCES

[1] Yan Z and Xu Y, "Data-Driven Load Frequency Control for Stochastic Power Systems: A Deep Reinforcement Learning Method With Continuous Action Search," *IEEE Trans. Power Syst.*, vol. 34, no. 2, pp. 1653-1656, 2019.

[2] Wang Y and Pedram M, "Model-Free Reinforcement Learning and Bayesian Classification in System-Level Power Management," *IEEE Trans. Comput.*, vol. 65, no. 12, pp. 3713-3726, 2016.

[3] Ge H, Song Y, Wu C, Ren J and Tan G, "Cooperative Deep Q-Learning With Q-Value Transfer for Multi-Intersection Signal Control," *IEEE Access*, vol. 7, pp. 40797-40809, 2019.

[4] Keerthisinghe C, Verbic G, Chapman A C, "A Fast Technique for Smart Home Management: ADP with Temporal Difference Learning," *IEEE Trans. Smart Grid*, vol. 9, no. 4, pp. 3291–3303, 2018.

[5] Huang H, Li F and Mishra Y, "Modeling Dynamic Demand Response Using Monte Carlo Simulation and Interval Mathematics for Boundary Estimation," *IEEE Trans. Smart Grid*, vol. 6, no. 6, pp. 2704–2713, 2015.

[6] Yu N P, Liu C C and Price J, "Evaluation of Market Rules Using a Multi-Agent System Method," *IEEE Trans. Power Syst.*, vol. 25, no. 1, pp. 470-479, 2010.

[7] Kebriaei H, Rahimi-Kian A and Ahmadabadi M N, "Model-Based and Learning-Based Decision Making in Incomplete Information Cournot Games: A State Estimation Approach," *IEEE Trans. Syst., Man, Cybern., Syst.*, vol. 45, no. 4, pp. 713–718, 2015.

[8] Peng Z, Hepeng L, Haibo H and Shuhui Li, "Dynamic Energy Management of a Microgrid using Approximate Dynamic Programming and Deep Recurrent Neural Network Learning," *IEEE Trans. Smart Grid*, vol. 10, no. 4, pp. 4435–4445, 2018.

[9] Kebriaei H, Ahmadabadi M N and Rahimi-Kian A, "Simultaneous State Estimation and Learning in Repeated Cournot Games," *Appl. Artif. Intell.*, vol. 28, no. 1, pp. 66-89, 2014.

[10] Iwata K, "Extending the Peak Bandwidth of Parameters for Softmax Selection in Reinforcement Learning," *IEEE Trans. Neural Netw. Learn. Syst.*, vol. 28, no. 8, pp. 1865-1877, 2017.

[11] Markus L and Magnus K, "Multiple Nash Equilibria in Electricity Markets with price-making Hydrothermal Producers," *IEEE Trans. Power Syst.*, vol. 34, no. 1, pp. 422-431, 2018.

[12] Savelli I, Giannitrapani A, Paoletti S and Vicino A, "An Optimization Model for the Electricity Market Clearing Problem with Uniform Purchase Price and Zonal Selling Prices," *IEEE Trans. Power Syst.*, vol. 33, no. 3, pp. 2864-2873, 2018.

[13] Dehghanpour K, Nehrir M H, Sheppard J W and N C.Kelly, "Agent-Based Modeling in Electrical Energy Markets Using Dynamic Bayesian Networks," *IEEE Trans. Power Syst.*, vol. 31, no. 6, pp. 4744-4754, 2016.